%% file: main.tex
\documentclass[conference]{IEEEtran}
\pdfoutput=1
\IEEEoverridecommandlockouts

\usepackage{graphicx}

\usepackage{xcolor}

\renewcommand{\arraystretch}{1.2}
\usepackage{multirow}
\usepackage{tablefootnote}

\usepackage{url}
\usepackage{svg}

\usepackage{tikz}

\usepackage{pifont}
\newcommand{\cmark}{\ding{51}}
\newcommand{\xmark}{{\color{red}\ding{55}}}

\def\BibTeX{{\rm B\kern-.05em{\sc i\kern-.025em b}\kern-.08em
    T\kern-.1667em\lower.7ex\hbox{E}\kern-.125emX}}

\begin{document}

\title{Enabling Performant and Secure EDA as a Service in Public Clouds Using Confidential Containers}

\author{Mengmei Ye \enspace Derren Dunn \enspace Daniele Buono \enspace Angelo Ruocco \\ Claudio Carvalho \enspace Tobin Feldman-Fitzthum \enspace Hubertus Franke \enspace James Bottomley*\thanks{*work done while at IBM Research.} \\IBM Research\\mye@ibm.com, dunnderr@us.ibm.com, dbuono@us.ibm.com,  ang@zurich.ibm.com, \\ \{cclaudio, tobin\}@ibm.com, \{frankeh, jejb\}@us.ibm.com
}

\maketitle
\thispagestyle{plain}
\pagestyle{plain}

\begin{abstract}
\input{abstract}
\end{abstract}


\input{intro}
\input{related-work}
\input{system}
\input{evaluation}

\input{RQs}
\input{conclusion}

\bibliographystyle{IEEEtran}

\end{document}

%% file: abstract.tex
Increasingly, business opportunities available to fabless design teams in the semiconductor industry far exceed those addressable with on-prem compute resources. An attractive option to capture these electronic design automation (EDA) design opportunities is through public cloud bursting. However, security concerns with public cloud bursting arise from having to protect process design kits, third party intellectual property, and new design data for semiconductor devices and chips.
One way to address security concerns for public cloud bursting is to leverage confidential containers for EDA workloads. Confidential containers  add zero trust computing elements to significantly reduce the probability of intellectual property escapes. A key concern that often follows security discussions is whether EDA workload performance will suffer with confidential computing. In this work we demonstrate a full set of EDA confidential containers and their deployment and characterize performance impacts of confidential elements of the flow including storage and networking. A complete end-to-end confidential container-based EDA workload exhibits 7.13\% and 2.05\% performance overheads over bare-metal container and VM based solutions, respectively. 

%% file: intro.tex
\section{Introduction} \label{sec:intro}

Cloud computing has the potential to significantly improve the resource availability of electronic design automation (EDA) in the semiconductor industry. The perception of infinite, ready-to-use compute power can provide a number of advantages to EDA workloads. It reduces both the need to plan in advance for compute capacity and the under-utilization of on-prem resources, decreasing time to market for chip design vendors~\cite{cloud-native-synopsys} and reducing the initial investments for smaller players and start-ups. However, one of the major challenges of offloading EDA workloads onto public cloud is maintaining a level of security comparable to on-prem deployments given that cloud infrastructure is shared with other tenants, and managed by a third party - the cloud service provider (CSP). 
This is of particular importance for EDA workloads, since they contain highly sensitive and valuable assets, such as intellectual properties (IPs) and process design kits (PDKs)~\cite{cloud-ready-synopsys}. A leak of EDA data can compromise competitiveness and even permanently damage brand reputation. While companies are (slowly) starting to exploit public cloud resources, the perceived risk of offloading EDA computations is still very high.

CSPs are offering several degrees of data protection. Most CSPs offer state-of-the-art ways to protect data at rest (when stored) and in motion (when sent over the network). However, this approach has historically presented two main security problems:
(1) When data reaches the compute unit (i.e. bare-metal server or, in most cases, virtual machines - VMs), data is decrypted. Any entity that is able to access the physical system running the workload, can exfiltrate data; and (2) There must be trust in the CSP: the provider is in most cases the only entity able to verify that data is really protected. It has access to the data itself and to the keys used to protect the data.

Confidential Computing is a fairly new technology that promises to remove such problems by additionally protecting data in use and allowing the user to verify the compute environments and maintain total control over encryption keys. This adds significant value to a use case such as EDA, where control over data is of paramount importance, and clients need to maintain a security model that is similar to current on-prem deployments.
 
With Confidential Computing, applications can be deployed into trusted execution environments (TEEs). The memory of a TEE is only accessible by a process running inside the TEE. Additionally, TEEs offer mechanisms to verify that the process is indeed running inside a TEE, and to attest to the content of the TEE. These are important mechanisms that allow a user to verify that the boundary of the TEE has not been compromised. By decrypting sensitive data retrieved from storage and networks only inside a verified TEE, we can guarantee that the CSP, or any attacker that gains access to CSP infrastructure, is unable to access sensitive data.~\cite{Azure-CC,Google-CC,AWS-CC,IBM-CC}.

As previously mentioned, storage and network encryption has been widely adopted by CSPs~\cite{Azure-network-enc,Google-network-enc,Google-storage-enc,IBM-storage-enc}. 
Historically, CSPs have focused more on allowing transparent workload integration; therefore, existing secure storage and network mechanisms mostly rely on the CSP hypervisor to conduct encryption/decryption. This  does not align with the security model of  Confidential Computing, where cloud hypervisors are outside the trust domain. A solution for secure storage and network that is compatible with the trust model of Confidential Computing is today left to be defined and implemented entirely by the user of Confidential Computing solutions.

EDA applications are historically deployed in on-prem clusters, and managed through job schedulers such as Slurm~\cite{slurm} and IBM Spectrum\textsuperscript{\textregistered} LSF\textsuperscript{\textregistered}~\cite{LSF}. As such, most current deployments of EDA applications in cloud environments use the same tools to provide a transparent experience for clients. While it is possible and in many cases beneficial to use job schedulers on public, private and hybrid cloud deployments, in this article we will assume that the EDA workload is deployed as a cloud-native application. Several vendors have already announced their intent to offer cloud-native versions of EDA tools, that would be able to leverage many of the tools available to deploy, administer and manage cloud-native applications. The same convenience applies here: we claim that by leveraging a cloud-native description of an EDA workload, we can introduce Confidential Computing aspects transparently without having to change the application itself, but only the orchestration around it.

In particular, we will leverage Confidential Containers (CoCo)~\cite{CoCo}, an open-source project that promises to transparently introduce Confidential Computing to a cloud-native application, with no changes required to the container images of the workload, and very minimal changes to the definition of the workload. The CoCo project is still in its infancy and has not tackled the issue of containers dealing with sensitive data stored on disk or communicated through the network. Architecting a solution for secure storage and network that is compatible with the trust model of Confidential Computing is one of the contributions of this paper. Additionally, even if the secure storage and network mechanisms adopted are in line with the security model of Confidential Computing, its performance implications with high-performance, large scale applications are still unclear.
More formally, in this paper, we aim to address the following research questions (RQs):

\begin{itemize}
    \item \textbf{RQ1 - Security}: How do we integrate secure storage and encrypted networks with Confidential Computing for cloud native EDA workloads? 
    \item \textbf{RQ2 - Performance}: What is the performance overhead introduced by Confidential Computing, secure storage, and encrypted network mechanisms? 
    \item \textbf{RQ3 - Automation}: Can we deploy end-to-end Confidential Computing automatically without making any modifications to the EDA workloads? 
\end{itemize}

To answer these RQs, we demonstrate a case study on a Kubernetes (K8s) cluster~\cite{k8s} that contains 28 bare-metal worker nodes (1736 CPU cores) with support for AMD Secure Encrypted Virtualization (SEV). In the K8s cluster, we evaluate Siemens Calibre\textsuperscript{\textregistered} optical proximity correction (OPC), a parallel distributed EDA workload, using CoCo. We add support for secure storage and encrypted networks in a generic way that does not depends on the specific workload. Based on our evaluation results, we further analyze and evaluate the remaining research challenges and opportunities to bridge the gap between EDA as a service and Confidential Computing. 

The rest of the paper is structured as follows. Section~\ref{sec:related-work} elaborates on the Siemens Calibre\textsuperscript{\textregistered} workloads, the state-of-the-art in cloud native EDA, and Confidential Computing. 
Section~\ref{sec:system} provides a system overview of our proposed framework. Section~\ref{sec:evaluation} elaborates on the evaluation results. 
Section~\ref{sec:RQs} provides answers and discussions to the RQs mentioned earlier. Section~\ref{sec:conclusion} concludes the paper. 

%% file: related-work.tex
\section{Background and Related Work} \label{sec:related-work}

\subsection{Siemens Calibre\textregistered ~OPC}
\label{subsec:calibre}
\begin{figure}[htp!]
\centerline{\includegraphics[width=0.9\columnwidth]{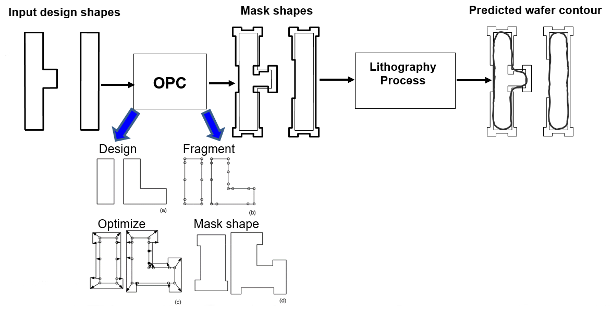}}
\caption{Optical Proximity Correction - shown is a representation of OPC.  Design shapes are read in and broken into vertices and edges. Next, simulations are performed to predict how the current shape will print on wafer for a particular lithography process.  Edges are then moved and new vertices or edges are introduced based upon the differences between a target shape and the simulated contour.  Finally, shapes are written to a file when either a cost function criteria is met or a maximum number of optimization iterations is reached.  }
\label{fig:OPC}
\end{figure}

In this work, we focus on a post-design application called Optical Proximity Correction (OPC)~\cite{calibre,calibre-opc}. OPC is a process in which design shapes are modified by optimization code to ensure that each shape will be transferred to wafer through  photolithography processes to within a predetermined tolerance~\cite{yan2007advances,kingsley2007advances}. In general, OPC runs begin by reading in a design database and partitioning shapes into vertices and edges as shown in Fig.~\ref{fig:OPC}.  Next, an iterative optimization process is run in which simulated contours for the targeted lithography process are compared with target shapes.  If a cost function criteria is not satisfied, edges are moved again until simulated contours match target shapes to within a predetermined criteria.  If the cost function is satisfied or a maximum number of iterations are reached, resultant shapes are written to a file to be merged later.

\begin{figure}[htp!]
\centerline{\includegraphics[width=0.6\columnwidth]{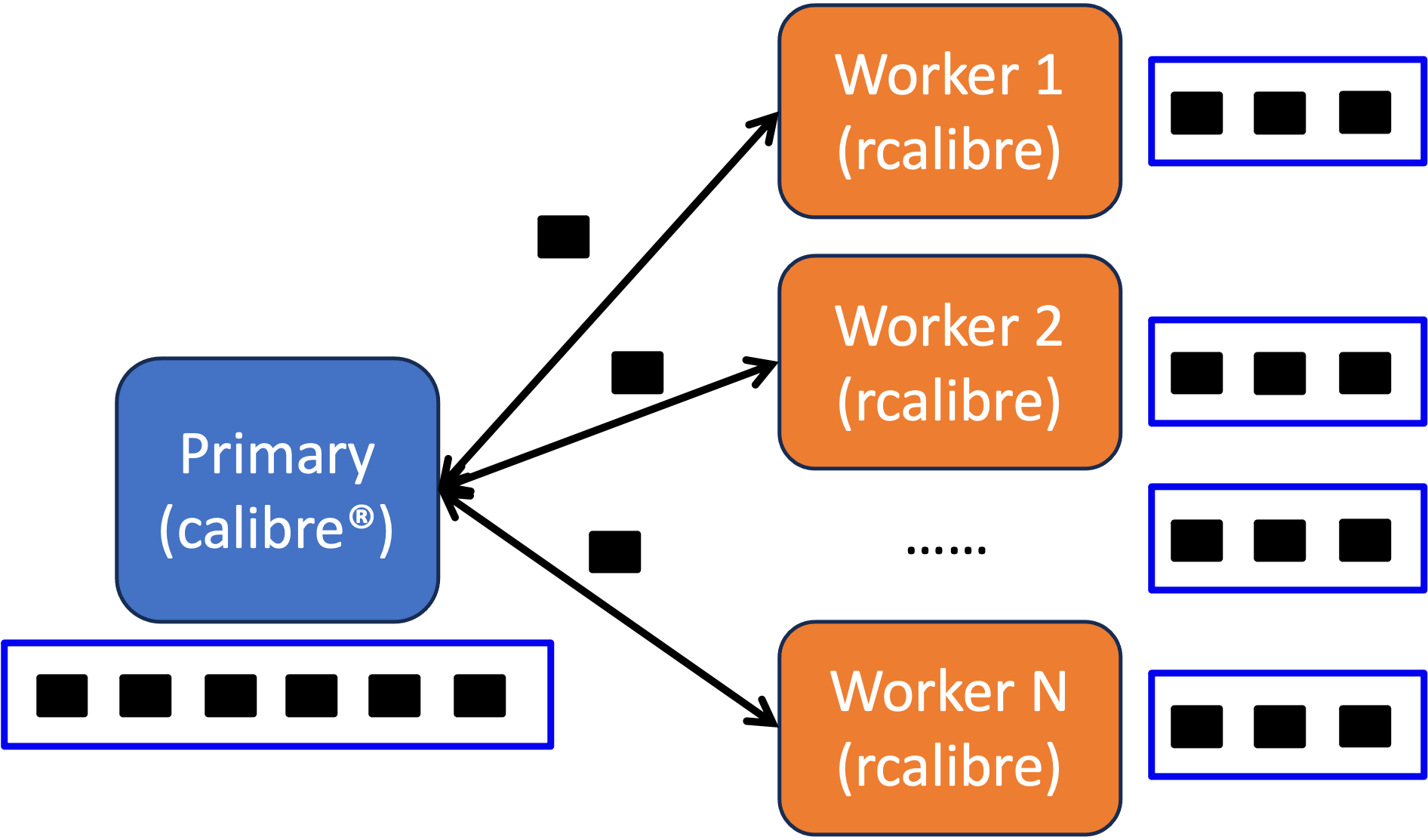}}
\caption{Overview of Siemens Calibre\textsuperscript{\textregistered} OPC as a distributed high performance computing workload. }
\label{fig:calibre}
\end{figure}

In practice, OPC is run as a traditional hub-and-spoke embarrassingly parallel application~\cite{spence2009computational} as shown in Fig.~\ref{fig:calibre}. Each OPC run is executed in two stages. First, a primary process reads in design layouts and partitions them into elements whose size is determined by the wavelength of light required by a target lithography process.  The primary process then creates a queue of elements that need to be optimized by OPC. Next, after a queue of elements is created, the primary process triggers the creation of worker processes that will process elements from the primary queue.  When the element queue is exhausted, the primary process then stitches together the results from the workers and writes out a merged file that can be used to create a photolithography mask. Typically, OPC runs are executed with thousands to tens of thousands of workers on Linux clusters.  

In this work, we chose to start with a distributed, compute intensive workload like OPC because it stresses many common elements of EDA workloads which include in CPU intensive operations, metadata operations to a shared file system, networking, and intensive input-output operations. We have discovered over several years of work that if we can run OPC efficiently in cloud environments, most other digital design tools can be run with little difficulty.

\subsection{Cloud Native EDA}
Currently, most OPC runs are executed on Linux clusters using either bare metal primary and worker nodes or in public clouds using VMs.  In general, it takes significant effort to provision, configure, and install required job management software, shared file systems, tune networks, and make connections to on-prem resources.  With this approach, it is possible to spend significant time and money provisioning and configuring suitable cloud infrastructure prior to executing a single useful run.  In addition, running in a multi-cloud environment compounds the complexity of infrastructure provisioning, configuration, and marshaling data from one cloud to another.
One mode of execution that reduces the complexity of provisioning and configuring infrastructure is to target Kubernetes job flows that leverage managed Kubernetes services offered by most public clouds.  By leveraging Kubernetes based flows, firms can store specific tool releases in centralized container registries and run anywhere with tools configured once in a container image.  In addition,  Kubernetes offers several cluster management features as part of a standard install like job autoscaling, storage services, automatic pod restarts that enable flexibility and ease of use.  As more and more users explore Kubernetes for applicable HPC workloads, 
 there will be an increased focus on methods and approaches to secure Kubernetes based workloads.
When thinking through how to migrate existing workloads to secure confidential containers workflows, one of the first questions that most groups will ask is what are the performance impacts of Kubernetes and of inserting confidential containers elements.  Previous authors have explored the impact of transitioning from bare metal on-prem systems to virtualized public cloud instances~\cite{hosny2021characterizing} which is an important first step.  In this work, we will show the impact of introducing elements of zero-trust computing into kubernetes based workflows for OPC.

\subsection{Confidential Computing}
There are different types of confidential computing techniques that deploy applications into TEEs/enclaves for various systems including x86, ARM, RISC-V, etc.~\cite{AMD-SEV, Intel-TDX,Intel-SGX, ARM-CCA, IBM-secure-execution, Keystone-Lee}.
Specifically for x86-based systems that our work targets, TEE technology primarily includes (1) process-based TEEs, such as Intel SGX~\cite{Intel-SGX} and (2) VM based TEEs, such as AMD SEV(-ES/-SNP)~\cite{AMD-SEV} and Intel TDX~\cite{Intel-TDX}. Process-based TEEs can achieve the smallest trusted computing base (TCB), where each application must be partitioned into a TEE and a non-TEE. The disadvantage is such approach requires to modify applications and causes significant performance overhead~\cite{meni2017exitlessSGX,miwa2023analyzing}. In contrast, VM-based TEEs deploy the entire VM (application plus operating system - OS) into an enclave. Although this increases the TCB size, it significantly improves convenience since it allows to keep the applications unchanged, and isolate customers from CSPs. Existing works that apply confidential computing to protect workloads are mainly based on confidential VMs or Intel SGX~\cite{meni2017exitlessSGX,akram2021performance,li2023IOperformance}.

\subsection{Confidential Containers}
Introducing support for Confidential Computing in a cloud-native application is not trivial. The main unit of computation with cloud-native applications is a container, which does not directly map to either process-based or VM-based TEEs. While there are works addressing process-based TEE with containers~\cite{Enclave-CC}, most of the research is focused on VM-based TEEs. Constellation by Edgeless Systems~\cite{Constellation} uses cluster-based TEEs and deploys the entire k8s cluster into a set of enclaves. This approach guarantees the highest degree of compatibility with cloud-native workloads and  transparent support. However, it makes verification/attestation of the content of the TEE difficult, and requires trusting the entire Kubernetes control plane, which is usually deployed by a CSP in managed offerings. A different, pod-centric approach is proposed by CoCo~\cite{CoCo}, where the TEE boundary is a pod. This still allows to deploy multiple containers within a single TEE, allowing resource sharing when required, but significantly reduces the size of the TEE, removes the need to trust the control plane, and makes attestation/verification easier, since the content of the VM is now only the set of containers within the pod - which can be verified by signing container images - and a very small, immutable OS used to bootstrap the pod.

\begin{figure}[htp!]
\centerline{\includegraphics[width=0.5\columnwidth]{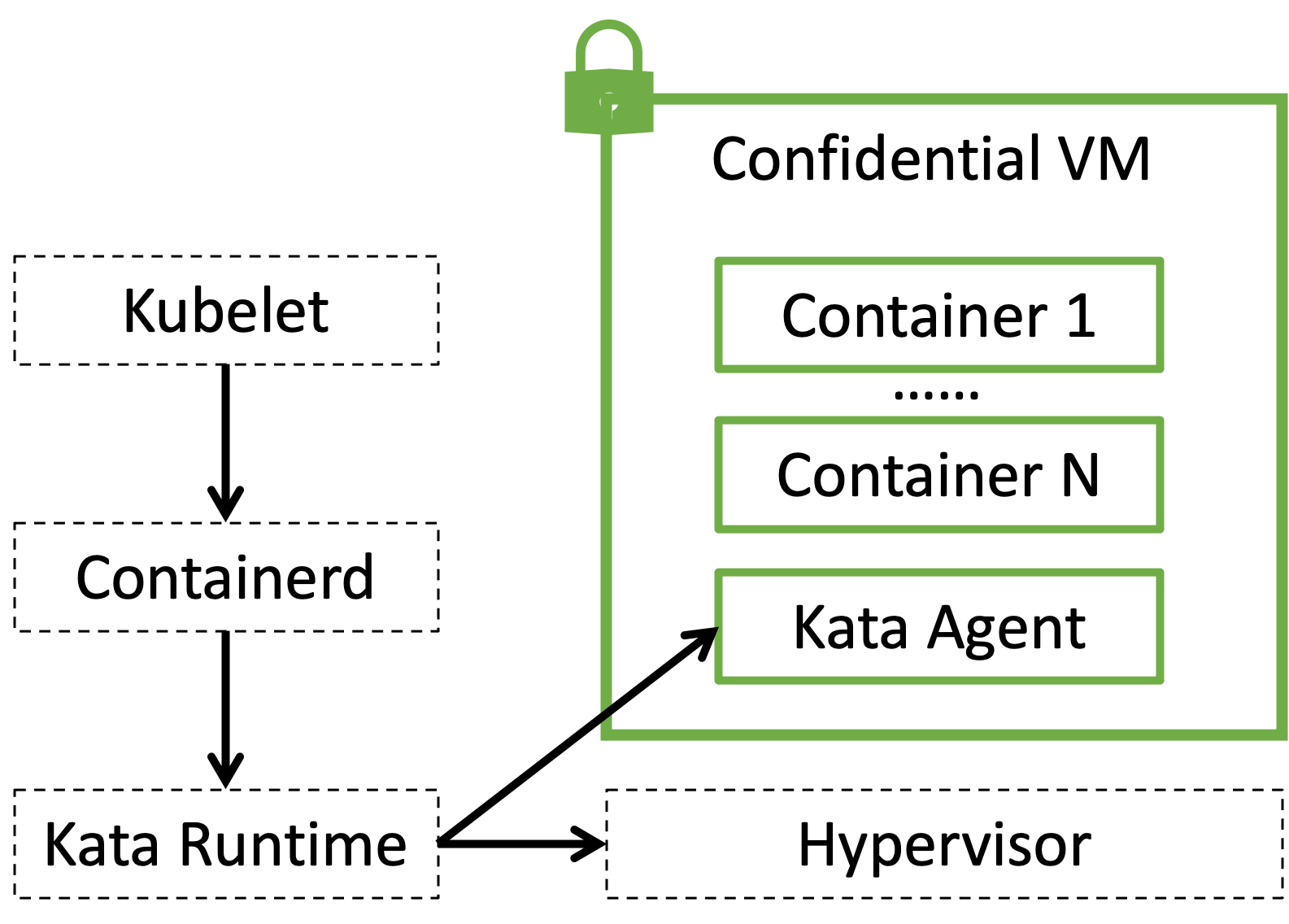}}
\caption{The components outside of the confidential VM are not trusted (black dashed boxes). The components in the confidential VM are encrypted (green solid-line boxes).}
\label{fig:CoCo}
\end{figure}

CoCo is based on Kata containers (Kata)~\cite{kata}. Kata aims at improving container isolation by deploying 
them into VMs, and introduces the concept of pod-level virtualization; however, there are several main differences between Kata and CoCo, given the different threat profile:
\begin{itemize}
    \item CoCo runs in confidential VMs, while Kata runs in regular unencrypted VMs.
    \item In Kata, most data is kept on the host, and shared with the VMs to increase sharing between pods. In CoCo, the host is not trusted, so the data,  including container images, must be kept inside the VM.
    \item CoCo and Kata use the same agent inside the VM to manage containers and processes. In CoCo, the agent must be attested, and the supported API calls highly restricted for security reasons. CoCo disables post-startup command execution into containers.
\end{itemize}
As shown in Fig.~\ref{fig:CoCo}, the containers and the Kata agent that manages the containers are deployed in the confidential VM, and the data is only decrypted inside of the VM. The rest of the components in the figure do not belong to the TEE, and therefore are not trusted. This includes the Kubelet that deploys pods into the node, Containerd that manages the container lifecycles, Kata Runtime that starts VMs and deploys containers into VMs, and the Hypervisor that manages the VMs.

To the best of our knowledge, this is the first work leveraging CoCo for an HPC-like workload at medium scale (28 worker nodes and more than 1700 CPU cores). Existing works on attacks and defenses~\cite{PwrLeak-Li,sidechannel-Li} in Confidential Computing are orthogonal to this work. 

%% file: system.tex
\section{System Framework} \label{sec:system}

\begin{figure}[htp!]
\centerline{\includegraphics[width=0.7\columnwidth]{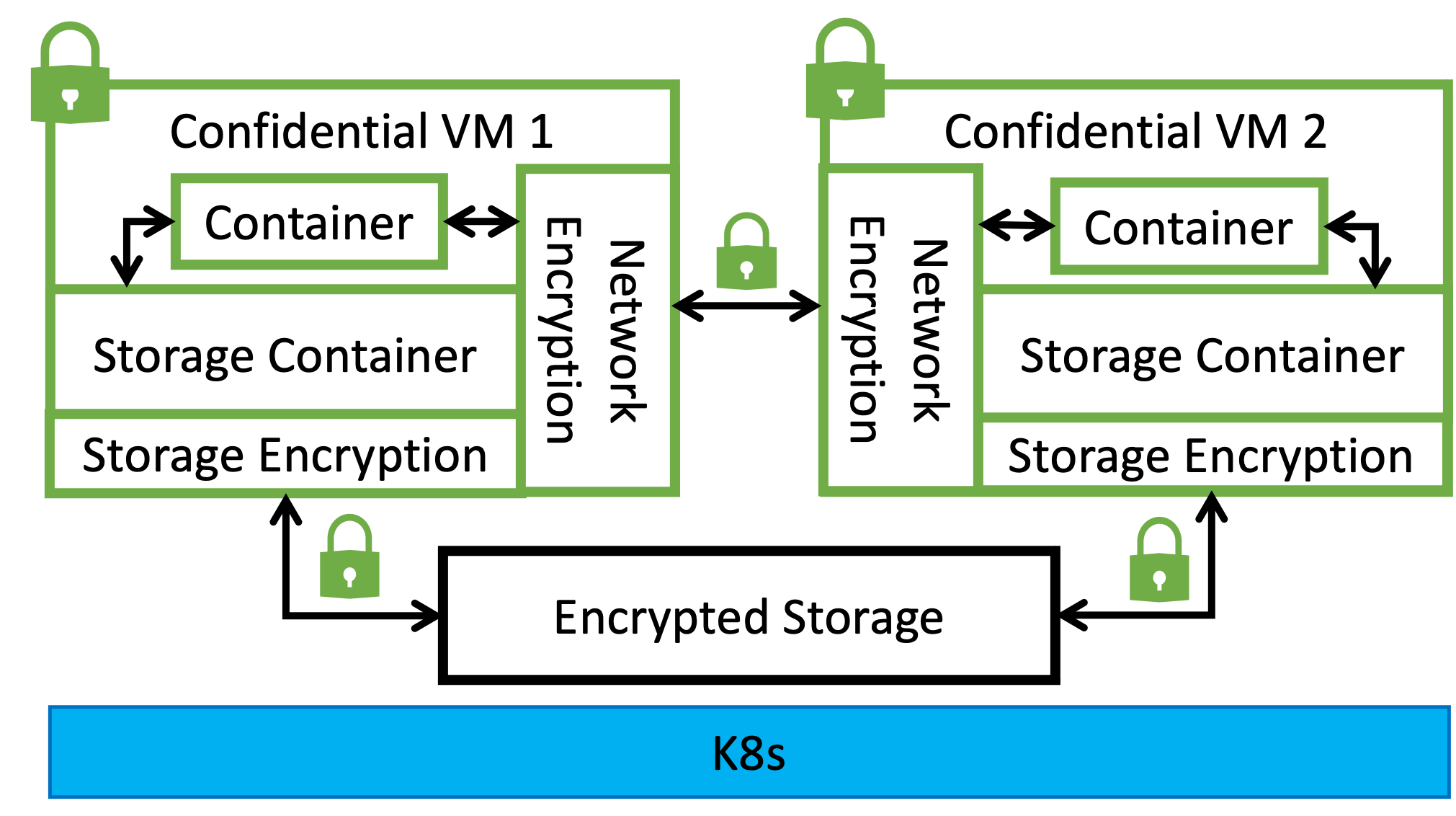}}
\caption{The system overview of CoCo with secure storage and encrypted networks.}
\label{fig:CoCo-system}
\end{figure}

Without Confidential Computing, the cloud native EDA workloads are deployed into classic or Kata containers in a K8s cluster. Here, the shared file system is mounted in the host and shared with the containers. In the following subsections, we will elaborate on how we leverage CoCo, and implement secure storage, and encrypted networks. The system overview for CoCo is shown in Fig.~\ref{fig:CoCo-system}. 

\subsection{Confidential Containers: Protect Data in Compute}

To protect data in compute, all workload pods are deployed into CoCo. The resource allocation (e.g., number of confidential containers, how many CPUs and memory resources allocated to each container) depends on the workload characteristics and available resources. The K8s control plane is not trusted. Data is only decrypted inside of confidential VMs. It is worth noting that the deployment with CoCo does not require any modifications to the workloads images. The deployment with CoCo can be configured with changes to pod specifications in yaml files. 

\subsection{Secure Storage: Protect Data at Rest}

EDA data, such as PDKs and derivative designs, is highly sensitive and must be protected securely. Classic encrypted storage mechanisms typically require the host to encrypt/decrypt data. To address this challenge, we leverage pod-level TEEs by deploying a sidecar in charge of transparently decrypting the data inside the TEE. We can do this either by leveraging Filesystems (FS) with native encryption, such as General Parallel File System (GPFS), Ceph File System (CephFS), or Network File System (NFS) with Kerberos, or by using a generic encryption layer on top of a non-encrypted filesystem. In any case, the sidecar will mount the file system inside the TEE and share it with the workload container.
Passwords or keys to encrypt/decrypt data can be securely transferred into confidential VMs via K8s Sealed Secret~\cite{sealed-secrets}. 

\subsection{Encrypted Networks: Protect Data in Motion}

When EDA workloads are deployed into multiple containers, the communication among the containers needs to be protected since it contains sensitive data, such as EDA design layouts.

Similarly to the secure storage solution, we deploy an additional sidecar container, in charge of encrypting the traffic between pods with mTLS. Such deployment guarantees that (1) data leaving the confidential VM is encrypted by the sidecar, and (2) only the pods that belong to the same workload can communicate with each other. 
As before, the keys and identities can be transferred using sealed secrets.

%% file: evaluation.tex
\section{Experimental Evaluation}
\label{sec:evaluation}

\subsection{Experimental Setup}

We deploy a K8s cluster on 29 AMD SEV-SNP bare-metal machines. Each of them has two AMD EPYC 7543 32-core processors. The K8s cluster contains 1 control plane and 28 worker nodes. Each of the worker nodes contains two pods, and each pod contains one container running the Siemens Calibre\textsuperscript{\textregistered} OPC. For the primary pod, we deploy 31 Calibre\textsuperscript{\textregistered} processes with $512$GB memory. For each of the worker pods, we deploy 31 rcalibre processes with the same amount of memory as the primary pod. In total, there are 56 pods and 1736 CPU cores.  

\begin{table*}[]
\centering
\caption{Comparison among different experimental setups.}
\label{tab:comparison}
\renewcommand{\arraystretch}{1.3}
\begin{tabular}{c|cccc}
\hline
                                         & \begin{tabular}[c]{@{}c@{}}Isolation\\ Boundary\end{tabular} & \begin{tabular}[c]{@{}c@{}}Shared Storage \\ Inside Isolation Boundary\end{tabular} & \begin{tabular}[c]{@{}c@{}}Encrypted \\ Shared Storage\end{tabular} & \begin{tabular}[c]{@{}c@{}}Encrypted \\ Network Traffic\end{tabular} \\ \hline
Experimental Baseline (K8s - Bare Metal) & Container                                                            & \xmark                                                                                    & \xmark                                                                    & \xmark                                                                     \\
Kata - Classic VM                        & VM                                                            & \xmark                                                                                    & \xmark                                                                    & \xmark                                                                     \\
CoCo - Confidential VM                   & TEE                                                            & \xmark                                                                                    & \xmark                                                                    & \xmark                                                                     \\
CoCo - Storage in Sidecar                & TEE                                                            & \cmark                                                                                    & \xmark                                                                    & \xmark                                                                     \\
CoCo - Encrypted Storage in Sidecar      & TEE                                                            & \cmark                                                                                    & \cmark                                                                    & \xmark                                                                     \\
CoCo - End-to-End                        & TEE                                                            & \cmark                                                                                    & \cmark                                                                    & \cmark                                                                     \\ \hline
\end{tabular}
\end{table*}

To determine the performance overhead of confidential computing when deploying the OPC workload, we conduct the experiments based on the following setups as shown in Table~\ref{tab:comparison}:
\begin{itemize}
    \item \emph{Experimental Baseline (K8s - Bare Metal)}: The workload runs in classic containers on a bare metal K8s cluster using an NFS persistent volume mounted in the host without any security protection (i.e., no CoCo, secure storage, or secure networks).
    \item \emph{Kata - Classic VM}: Unlike the experimental baseline where the workload runs in classic containers, this setup deploys the workload with Kata containers where each pod is encapsulated in a VM. 
    \item \emph{CoCo - Confidential VM}: Compared to the Kata setup, this CoCo setup runs the workload with confidential containers where each pod is deployed inside its own confidential (AMD SEV) VM. The NFS persistent volume is still mounted in the host. 
    \item \emph{CoCo - Storage in Sidecar}: Unlike the CoCo setup that uses host-mounted NFS, this setup mounts the (unencrypted) file system in each confidential VM using a sidecar. 
    \item \emph{CoCo - Encrypted Storage in Sidecar}: Building on the previous setup, this configuration adds an encryption layer (i.e., fuse-based gocryptfs~\cite{gocryptfs}) to the persistent storage, which runs inside of the confidential VM. 
    \item \emph{CoCo - End-to-End}: Building on the previous setup, this configuration adds mTLS-encrypted networking by leveraging Istio Envoy sidecars~\cite{envoy} executing inside of the confidential VM. This configuration/setup provides a complete CoCo-based confidential end-to-end framework with data protection in compute, at rest, and in motion. 
\end{itemize}

\subsection{Performance Evaluation}
\label{subsec:perf-eval}
\begin{figure}[htp!]
\centerline{\includegraphics[width=1.0\columnwidth]{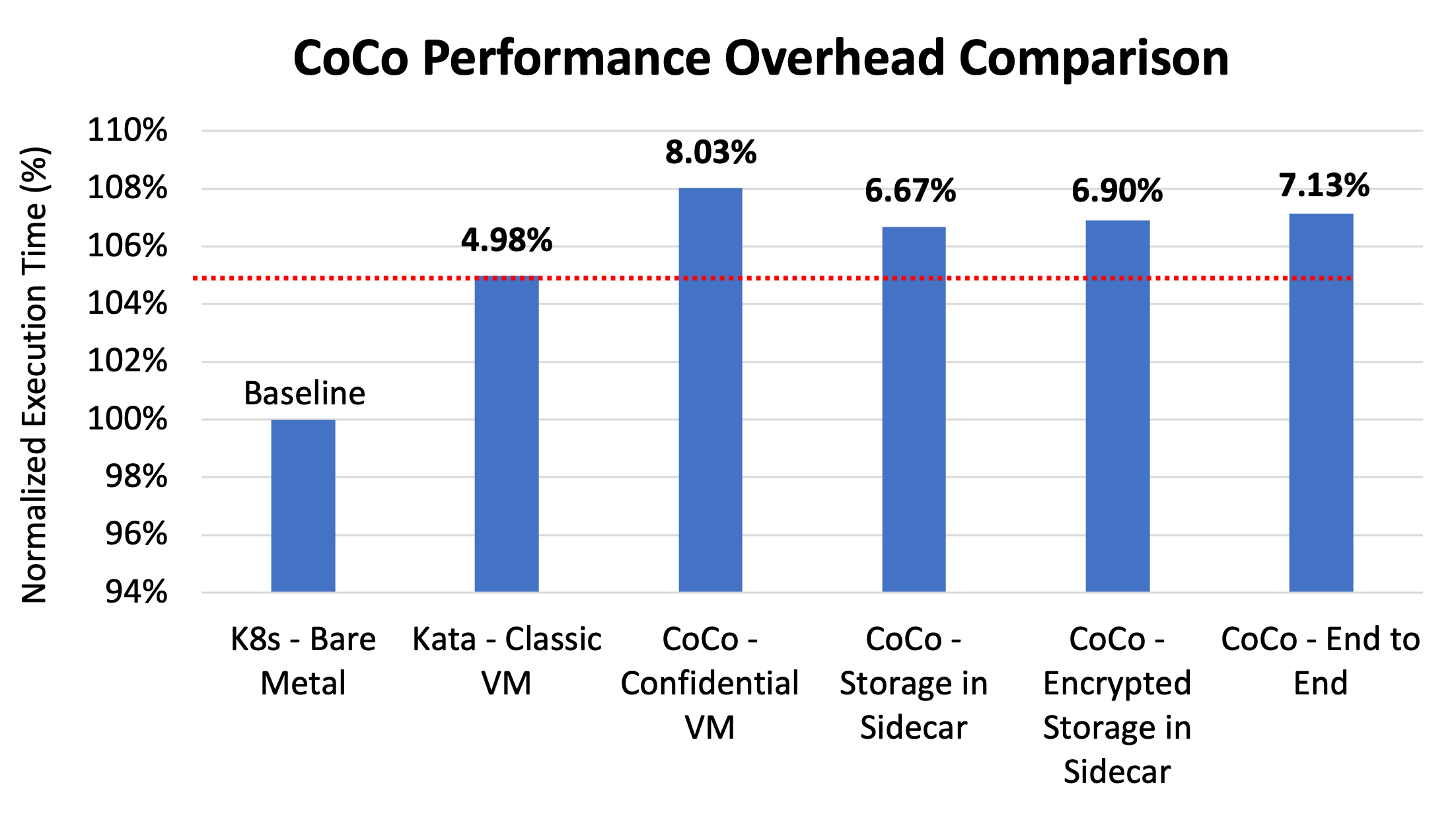}}
\caption{Performance overhead of k8s Bare Metal, Kata - Classic VM, and various configurations of CoCo.}
\label{fig:coco-overhead}
\end{figure}

In Fig.~\ref{fig:coco-overhead} we measure the overhead of each experiment, normalized to the baseline (K8s-BareMetal). The results show that Kata overhead is 4.98\% and CoCo overhead is 8.03\% compared with the k8s baseline. Moving the shared storage in the sidecar allows us to gain several benefits. First, it moves closer to the CoCo to protect the shared storage. Second, it eliminates the overhead of another virtualization and sharing later (storage mounted on the host needs to be then shared with each confidential VM). Third, it improves performance and reduces the overhead from 8.03\% to 6.67\%.

By adding the encryption layer on top of the storage sidecar, the total overhead is 6.9\% as shown in Fig.~\ref{fig:coco-overhead}. Furthermore, when we add network encryption to the Sidecar setup the results show that the CoCo - End-to-End with secure storage and secure networks adds 7.13\% overhead compared to the baseline K8s - Bare Metal. 

\begin{figure}[htp!]
\centerline{\includegraphics[width=0.8\columnwidth]{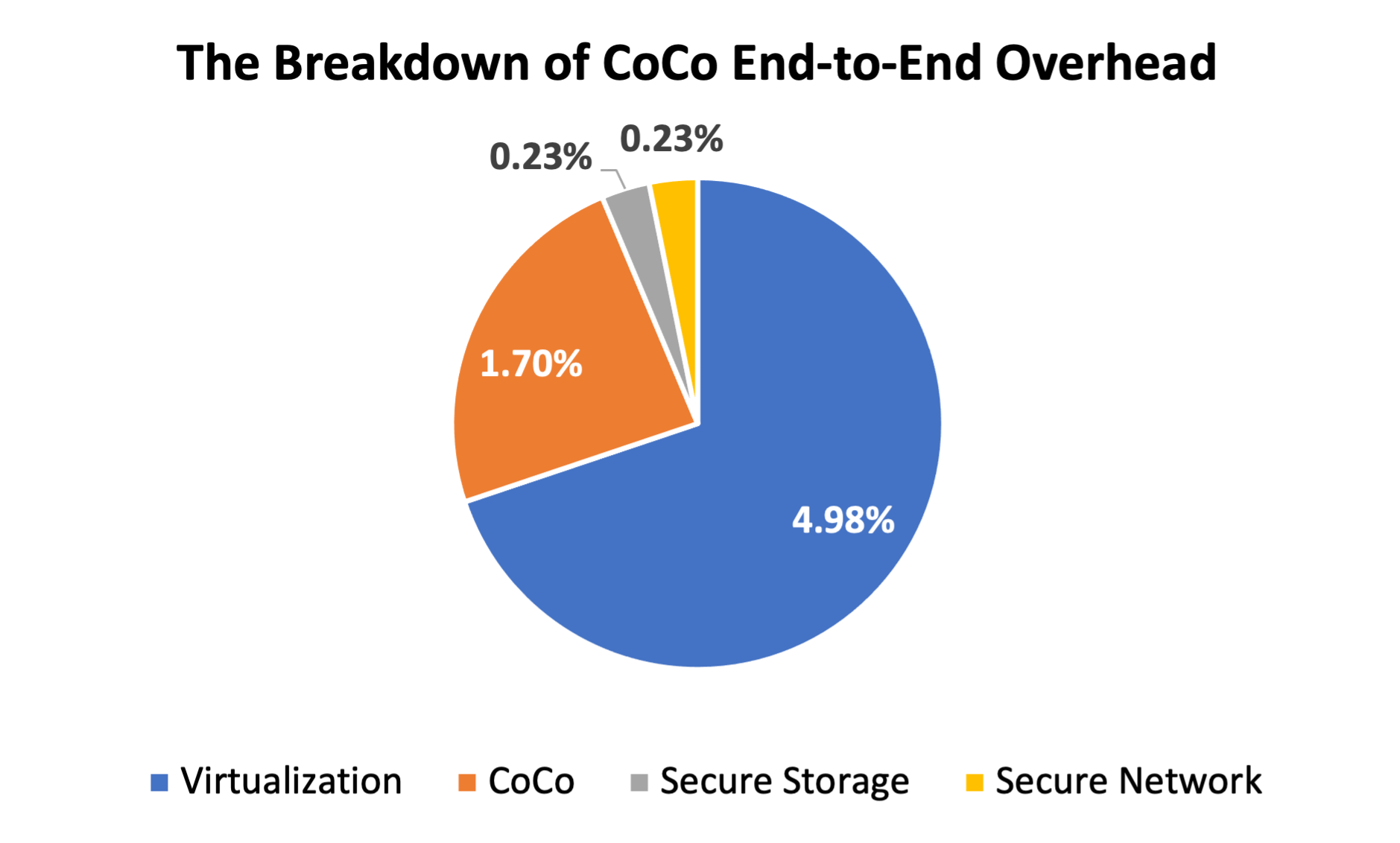}}
\caption{Performance overhead of each component in the proposed framework. }
\label{fig:coco-end2end}
\end{figure}

These experiments allow us to break down the performance overhead of each component of the proposed framework. We notice that virtualization consumes 4.98\% when the containers are deployed in VMs. Data protection in compute (CoCo), at rest (secure storage), and in motion (encrypted networks) consume 1.70\%, 0.23\%, and 0.23\%, respectively. It is worth noting that here the comparison baseline is K8s - Bare Metal. Considering most of the cloud workloads are deployed in VMs, the comparison baseline could also be considered a classic VM.
In this case, with the consideration of Kata-Classic VM as the alternative baseline, shown in Fig.~\ref{fig:coco-overhead} as the red dotted line, the overheads are 2.91\%, 1.62\%, 1.83\%, and 2.05\% for ``CoCo - Confidential VM'', ``CoCo - Storage in Sidecar'', ``CoCo - Encrypted Storage in Sidecar'', and ``CoCo - End to End'', respectively.

%% file: RQs.tex
\section{Research Questions}
\label{sec:RQs}

\subsection{RQ1 - Security}

Our proposed framework (Fig.~\ref{fig:CoCo-system}) leverages Confidential Containers, as well as secure storage and encrypted network sidecars, which achieves end-to-end Confidential Computing and protects data in compute, at rest and in motion. Every component in the confidential VMs (including injected storage and network sidecars) can be attested/measured before/after they are booted. Our proposed framework does not modify the booting process of confidential VMs/containers and thus it does not require any special/modified attestation/measurement mechanisms. Existing attestation/measurement mechanisms, such as CoCo attestation ``background check'' model~\cite{attestation}, can be applied to this framework. We will evaluate the attestation and measurement as our future work.

In summary, the confidential computing techniques remove the the CSPs from the trust domain and thus reduce the TCB size, and hence the attack surfaces. Security bugs in CoCo or attacks that TEEs are vulnerable to, such as side channel attacks~\cite{feldtkeller2023challenges}, are out of scope for this paper. 

\subsection{RQ2 - Performance}

Based on our performance evaluation in Section~\ref{subsec:perf-eval}, our proposed framework achieves performant and secure EDA as a service, where the end-to-end confidential computing only causes an around 7\% increase in execution time at a moderate scale (i.e., 28 worker nodes and more than 1700 CPU cores). There are also improvements we can contribute in the future. For example, in the CoCo - End-to-End setup, most of the performance overhead is from virtualization of classic VMs, which introduces an almost 5\% increase in execution time. Improving the overhead of virtualization will not only reduce the overhead of confidential containers but also reduce the overhead for XaaS (anything as a service) since most of the cloud applications run inside of VMs. Furthermore, the CoCo overhead can be improved with the future generations of VM-based TEEs or future CoCo releases. In the performance evaluation shown in Section~\ref{sec:evaluation}, we utilize CoCo v0.7.0 release and deploy it in our AMD SEV-SNP bare-metal nodes. CoCo is a rapidly evolving project, and thus performance results can be varied in future releases. We believe the analysis and evaluation in this paper will be beneficial for the future research and development of confidential computing for EDA as a service for both the confidential computing and cloud-native EDA communities. 

\subsection{RQ3 - Automation}

In the proposed framework there are no modifications required to the workloads due to the following reasons. First, CoCo leverages VM-based TEEs that can deploy the entire containerized workloads into the confidential/encrypted VMs directly. Second, the secure storage and network mechanisms leverage the sidecar technology and are deployed aside of the containers that run workloads in the confidential VMs. The configurations of these mechanisms can be customized in the yaml files before they are deployed in the system. In the experiments, we leverage gocryptfs as a generic sidecar for the encrypted file system layer. It is also applicable to deploy bring-your-own-sidecar for customized, non open-source solutions (e.g., GPFS). Similarly, in encrypted networks, we apply Envoy sidecars for mTLS encrypted network traffic, where Envoy can be replaced by other network encryption mechanisms. 

In the future, we will apply our proposed framework to more workloads, such as generic high performance computing (HPC) workloads and artificial intelligence (AI) workloads. In addition, we will extend the confidential computing to a larger scale, as well as considering accelerators. 

%% file: conclusion.tex
\section{Conclusion} \label{sec:conclusion}

In this paper, we address existing challenges and demonstrate a proof of concept for performant and secure EDA as a service by leveraging confidential computing techniques. In the experiments, we employ a Siemens Calibre\textsuperscript{\textregistered} OPC workload in confidential containers with storage and network encryption. The evaluation results show that confidential computing adds around 7\% performance overhead compared to running the workload in the unprotected containers in a bare-metal K8s cluster. We also analyze each portion of performance overhead and observe that virtualization causes around 5\% overhead and encryption causes around 2\% overhead. Last but not least, we discuss future research opportunities for performant and secure EDA as a service.

%% file: main.bbl
\begin{thebibliography}{10}
\providecommand{\url}[1]{#1}
\csname url@samestyle\endcsname
\providecommand{\newblock}{\relax}
\providecommand{\bibinfo}[2]{#2}
\providecommand{\BIBentrySTDinterwordspacing}{\spaceskip=0pt\relax}
\providecommand{\BIBentryALTinterwordstretchfactor}{4}
\providecommand{\BIBentryALTinterwordspacing}{\spaceskip=\fontdimen2\font plus
\BIBentryALTinterwordstretchfactor\fontdimen3\font minus
  \fontdimen4\font\relax}
\providecommand{\BIBforeignlanguage}[2]{{%
\expandafter\ifx\csname l@#1\endcsname\relax
\typeout{** WARNING: IEEEtran.bst: No hyphenation pattern has been}%
\typeout{** loaded for the language `#1'. Using the pattern for}%
\typeout{** the default language instead.}%
\else
\language=\csname l@#1\endcsname
\fi
#2}}
\providecommand{\BIBdecl}{\relax}
\BIBdecl

\bibitem{cloud-native-synopsys}
G.~Singh, ``What does cloud-native mean for chip designers?'' 2023,
  \url{https://www.synopsys.com/cloud/insights/what-does-cloud-native-mean.html}.

\bibitem{cloud-ready-synopsys}
``Is the {EDA} industry ready for cloud computing?'' 2023,
  \url{https://www.synopsys.com/cloud/insights/cloud-computing.html}.

\bibitem{Azure-CC}
``Azure confidential computing,'' 2023,
  \url{https://azure.microsoft.com/en-us/solutions/confidential-compute/}.

\bibitem{Google-CC}
``Confidential computing - {Google Cloud},'' 2023,
  \url{https://cloud.google.com/confidential-computing}.

\bibitem{AWS-CC}
D.~Brown, ``Confidential computing: an {AWS} perspective,'' 2021,
  \url{https://aws.amazon.com/blogs/security/confidential-computing-an-aws-perspective/}.

\bibitem{IBM-CC}
``Confidential computing on {IBM Cloud},'' 2023,
  \url{https://www.ibm.com/cloud/confidential-computing}.

\bibitem{Azure-network-enc}
``What is {Azure} virtual network encryption? (preview),'' 2023,
  \url{https://learn.microsoft.com/en-us/azure/virtual-network/virtual-network-encryption-overview}.

\bibitem{Google-network-enc}
``Encryption in transit,'' 2023,
  \url{https://cloud.google.com/docs/security/encryption-in-transit}.

\bibitem{Google-storage-enc}
``Default encryption at rest,'' 2023,
  \url{https://cloud.google.com/docs/security/encryption/default-encryption}.

\bibitem{IBM-storage-enc}
``{IBM} security guardium data encryption,'' 2023,
  \url{https://www.ibm.com/products/guardium-data-encryption}.

\bibitem{slurm}
``Slurm - workload manager,''
  \url{https://slurm.schedmd.com/documentation.html}.

\bibitem{LSF}
``{IBM} {Spectrum} {LSF} suites,''
  \url{https://www.ibm.com/products/hpc-workload-management}.

\bibitem{CoCo}
``Confidential containers,'' 2023,
  \url{https://github.com/confidential-containers}.

\bibitem{k8s}
``Kubernetes,'' 2023, \url{https://kubernetes.io}.

\bibitem{calibre}
``Siemens Calibre curvilinear solutions,''
  \url{https://eda.sw.siemens.com/en-US/ic/calibre-manufacturing/curvilinear-data-preparation/}.

\bibitem{calibre-opc}
``Siemens Calibre nmclopc,''
  \url{https://eda.sw.siemens.com/en-US/ic/calibre-manufacturing/curvilinear-data-preparation/nmclopc/}.

\bibitem{yan2007advances}
X.~Yan, Z.~Shi, Y.~Chen, and Q.~Chen, ``Advances in opc technology and
  development of zopc tool,'' in \emph{Quantum Optics, Optical Data Storage,
  and Advanced Microlithography}, vol. 6827, 2007, pp. 390--403.

\bibitem{kingsley2007advances}
T.~Kingsley, J.~Sturtevant, S.~McPherson, and M.~Sexton, ``Advances in compute
  hardware platforms for computational lithography,'' in \emph{Optical
  Microlithography XX}, vol. 6520, 2007, pp. 482--497.

\bibitem{spence2009computational}
C.~Spence and S.~Goad, ``Computational requirements for opc,'' in \emph{Design
  for Manufacturability through Design-Process Integration III}, vol. 7275,
  2009, pp. 230--238.

\bibitem{hosny2021characterizing}
A.~Hosny and S.~Reda, ``Characterizing and optimizing eda flows for the
  cloud,'' \emph{IEEE Transactions on Computer-Aided Design of Integrated
  Circuits and Systems}, pp. 3040--3051, 2021.

\bibitem{AMD-SEV}
``{AMD} secure encrypted virtualization,'' 2020,
  \url{https://www.amd.com/en/processors/amd-secure-encrypted-virtualization}.

\bibitem{Intel-TDX}
``Intel trust domain extensions,'' 2023,
  \url{https://www.intel.com/content/www/us/en/developer/articles/technical/intel-trust-domain-extensions.html}.

\bibitem{Intel-SGX}
``Intel software guard extensions,'' 2023,
  \url{https://www.intel.com/content/www/us/en/developer/tools/software-guard-extensions/overview.html}.

\bibitem{ARM-CCA}
``{ARM} confidential compute architecture,'' 2023,
  \url{https://www.arm.com/architecture/security-features/arm-confidential-compute-architecture}.

\bibitem{IBM-secure-execution}
``Introducing {IBM} secure execution for {Linux},'' 2022,
  \url{https://www.ibm.com/docs/en/linux-on-systems?topic=virtualization-secure-execution}.

\bibitem{Keystone-Lee}
D.~Lee, D.~Kohlbrenner, S.~Shinde, K.~Asanovi\'{c}, and D.~Song, ``Keystone: An
  open framework for architecting trusted execution environments,'' pp. 1--16,
  2020.

\bibitem{meni2017exitlessSGX}
M.~Orenbach, P.~Lifshits, M.~Minkin, and M.~Silberstein, ``Eleos: Exitless {OS}
  services for {SGX} enclaves,'' in \emph{European Conference on Computer
  Systems (EuroSys)}, 2017, p. 238–253.

\bibitem{miwa2023analyzing}
S.~Miwa and S.~Matsuo, ``Analyzing the performance impact of {HPC} workloads
  with {Gramine+SGX} on 3rd generation {Xeon} scalable processors,'' in
  \emph{Workshops of The International Conference on High Performance
  Computing, Network, Storage, and Analysis (SC-W 2023)}, 2023, pp. 1849--1858.

\bibitem{akram2021performance}
A.~Akram, A.~Giannakou, V.~Akella, J.~Lowe-Power, and S.~Peisert, ``Performance
  analysis of scientific computing workloads on general purpose {TEEs},'' in
  \emph{IEEE International Parallel and Distributed Processing Symposium
  (IPDPS)}, 2021, pp. 1066--1076.

\bibitem{li2023IOperformance}
D.~Li, Z.~Mi, C.~Ji, Y.~Tan, B.~Zang, H.~Guan, and H.~Chen, ``Bifrost: Analysis
  and optimization of network {I/O} tax in confidential virtual machines,'' in
  \emph{USENIX Annual Technical Conference (USENIX ATC)}, 2023, pp. 1--15.

\bibitem{Enclave-CC}
``Enclave-cc,'' \url{https://github.com/confidential-containers/enclave-cc}.

\bibitem{Constellation}
``Constellation,'' 2023,
  \url{https://www.edgeless.systems/products/constellation/}.

\bibitem{kata}
``Kata containers: the speed of containers, the security of {VMs},'' 2023,
  \url{https://katacontainers.io}.

\bibitem{PwrLeak-Li}
W.~Wang, M.~Li, Y.~Zhang, and Z.~Lin, ``Pwrleak: Exploiting power reporting interface for side-channel attacks on {AMD} {SEV},'' in \emph{Detection
  of Intrusions and Malware, and Vulnerability Assessment(DIMVA)}, 2023, p.
  46–66.

\bibitem{sidechannel-Li}
M.~Li, L.~Wilke, J.~Wichelmann, T.~Eisenbarth, R.~Teodorescu, and Y.~Zhang, ``A
  systematic look at ciphertext side channels on amd sev-snp,'' in \emph{IEEE
  Symposium on Security and Privacy (SP)}, 2022, pp. 337--351.

\bibitem{sealed-secrets}
``Kubernetes sealed secret,''
  \url{https://github.com/confidential-containers/guest-components/blob/main/confidential-data-hub/docs/SEALED_SECRET.md}.

\bibitem{gocryptfs}
``Gocryptfs - simple. secure. fast,'' \url{https://nuetzlich.net/gocryptfs/}.

\bibitem{envoy}
``Double proxy (with {mTLS} encryption),''
  \url{https://www.envoyproxy.io/docs/envoy/latest/start/sandboxes/double-proxy}.

\bibitem{attestation}
P.~Banerjee and S.~Ortiz, ``Understanding the confidential containers
  attestation flow,'' 2022,
  \url{https://www.redhat.com/en/blog/understanding-confidential-containers-attestation-flow}.

\bibitem{feldtkeller2023challenges}
J.~Feldtkeller, P.~Sasdrich, and T.~G{\"u}neysu, ``Challenges and opportunities
  of security-aware eda,'' \emph{ACM Transactions on Embedded Computing
  Systems}, pp. 1--34, 2023.

\end{thebibliography}
